\documentclass[preprint,aps,pre,amsmath,amssymb,amsfonts]{revtex4}

\def\Cb{{\Bbb C}}
\def\be{\begin{equation}}
\def\ee{\end{equation}}
\def\ba{\begin{array}}
\def\ea{\end{array}}
\def\Cb{\ \hbox{\vrule width 0.6pt height 6.5pt depth 0pt
              \hskip -3.2 pt} C}
\def\Rb{{I\!\! R}}
\def\qed{\leavevmode\unskip\penalty9999 \hbox{}\nobreak\hfill
     \quad\hbox{\leavevmode  \hbox to.77778em{%
               \hfil\vrule   \vbox to.675em%
               {\hrule width.6em\vfil\hrule}\vrule\hfil}}
     \par\vskip3pt}
\newtheorem{theorem}{Theorem}

\newtheorem{cor}[theorem]{Corollary}

\begin{document}

\title{On Locality of Schmidt-Correlated States}


\author{Ming-Jing Zhao}
\author{Shao-Ming Fei}

\affiliation{
{School of Mathematical Sciences, Capital Normal
University, Beijing 100048,
China}}

\begin{abstract}
We review some results on the equivalence of quantum states under local unitary
transformations (LUT). In particular, the classification of two-qubit Schmidt correlated (SC)
states under LUT is investigated.  By presenting the standard form of quantum states under LUT,
the sufficient and necessary conditions of whether two different SC states are local unitary equivalent
are provided. The correlations of SC states are also discussed.
\end{abstract}

\maketitle

Quantum entanglement has been extensively investigated
as a key physical resource in quantum information processing
such as quantum cryptography, quantum
teleportation, dense coding, quantum error correction, quantum repeater,
remote state preparation and quantum computation \cite{nielsen}.
One way to gain insight into the entanglement is to consider the interconvertibility
between quantum states. For two  given states $|\psi\rangle$ and
$|\phi\rangle$, the question is whether or not $|\psi\rangle$ can be
transformed into $|\phi\rangle$ by local operations \cite{horodeckirmp}.
A particular interesting kind of local operations is the local unitary (LU) transformation.
Since the properties of entanglement for bipartite and multipartite quantum systems
remain invariant under local unitary transformations on the individual subsystems,
it has been an important problem to classify quantum states under local unitary
transformations.

An n-partite pure state $|\psi\rangle$ is said to be LU equivalent to $|\phi\rangle$ if there exist
local unitary operators $U_1, \cdots, U_n$ such that $|\psi\rangle =
U_1\otimes \cdots \otimes U_n |\phi\rangle$. Two
n-partite mixed states $\rho_1$ and $\rho_2$ are called LU
equivalent if there exist local unitary operators $U_1, \cdots, U_n$
such that $\rho_1 = U_1\otimes \cdots \otimes U_n \rho_2
U_1^\dagger\otimes \cdots \otimes U_n^\dagger$.

One way to deal with the problem is to find the
complete set of invariants of local unitary transformations. Two
states are equivalent under local unitary transformations if and
only if they have the same values of all these invariants. The
method developed in \cite{Rains,Grassl}, in principle, allows one
to compute all the invariants of local unitary transformations,
though in general it is not operational.

The problem of LU equivalences for arbitrary dimensional bipartite pure states,
two-qubit mixed states and n-qubit pure states have been also solved completely.
We first consider the case of pure states. Let $H$ be an
$N$-dimensional complex Hilbert space, with $\vert i\rangle$, $i=1,...,N$,
as an orthonormal basis. A general pure state on $H\otimes H$ is of the form
\begin{equation}\label{mmm}
\vert\Psi\rangle=\sum_{i,j=1}^N a_{ij}\vert i\rangle \otimes
\vert j\rangle,~~~~~~a_{ij}\in\Cb
\end{equation}
with the normalization $\sum_{i,j=1}^N
a_{ij}\,a_{ij}^\ast=1$ ($\ast$ denoting complex conjugation).
A quantity is called an invariant associated with the
state $\vert\Psi\rangle$ if it is invariant under all local
unitary transformations, i.e. all maps of the form $U_1\otimes U_2$
from $H \otimes H$ to
itself, where $U_1$ and $U_2$ are unitary transformations on the Hilbert space
$H$. Let $A$ denote the matrix given by $(A)_{ij}=a_{ij}$. The following quantities are known to be
invariants associated with the state $\vert\Psi\rangle$ given by (\ref{mmm}),
see \cite{5678}:
\begin{equation}\label{I}
I_\alpha=Tr(AA^\dag)^{\alpha},~~~~~~~~~~~\alpha=1,...,N;
\end{equation}
(with $A^\dag$ the adjoint of the matrix $A$). It can be shown that two
pure bipartite states are LU equivalent if and only if they
have the same values of these invariants.

In \cite{Makhlin} the invariants for general two-qubit systems are studied and a
complete set of 18 polynomial invariants is presented.
It has been shown that these 18 invariants
are sufficient to guarantee that two two-qubit states are equivalent
under local unitary transformations, and lack of anyone of these
18 invariants would result in incompleteness of the set of invariants.
In \cite{Kraus} it has been shown that arbitrary n-qubit pure
quantum states are LU equivalent if and only if their computable
standard forms coincide.

Nevertheless, to determine
whether two general states are locally equivalent is still an challenging problem.
We have results only for some special classes of quantum pure or mixed states.
In \cite{linden,sud} three-qubit mixed states are discussed according to LU invariants.
In \cite{generic,sungeneric} a complete set of invariants is presented for
bipartite generic mixed states. In \cite{goswami} a complete set
of invariants under local unitary transformations is presented for
rank-2 and multiplicity free mixed states. In \cite{tri} the case
of tripartite is studied in detail and a complete set of
invariants is presented for a class of pure states.

Besides LU invariant approach, in \cite{jing-eq} the matrix tensor product approach
to the equivalence of quantum states under local unitary transformations
has been introduced. If two bipartite density matrices $\rho$ and $\rho^\prime$ in
$H_1\otimes H_2$, with dimensions $M$ and $N$ ($M\geq N$) respectively, are equivalent
under local unitary transformations, they must have the same set of
eigenvalues $\lambda_i$, $i=1,...,NM$. Let $X$ and $Y$ be the unitary
matrices that diagonalize $\rho$ and $\rho^\prime$ respectively,
\be\label{xy}
\rho=X\Lambda X^\dag,~~~~~\rho^\prime=Y\Lambda Y^\dag,
\ee
where $\Lambda=diag(\lambda_1,\lambda_2,...,\lambda_{MN})$.
Let $G$ be the fixed point unitary subgroup
associated with $\rho$, i.e. $g\rho=\rho g$ for all $g\in G$.
If we define that a matrix $V$ on $H_1\otimes H_2$ is tensor
decomposable if it can be written as $V=V_1\otimes V_2$ for
$V_1\in End(H_1)$, $V_2\in End(H_2)$, then $\rho^\prime$ is equivalent to $\rho$
under local unitary transformations if and only if the coset
$GXY^\dag$ contains a unitary tensor decomposable matrix.

Let $\rho$ and $\rho^\prime$ be two density matrices
with orthonormal unitary matrices $X$ and $Y$ as given in (\ref{xy}).
If $\rho$ and $\rho^\prime$ are not degenerate,
they are equivalent under local unitary transformations if and only if the set of
matrices $XDY^\dag$, $D=diag(e^{i\theta_1},\,e^{i\theta_2},...,e^{i\theta_{MN}})$,
contains a unitary tensor decomposable element for some $\theta_i\in \Rb$.
This conclusion presents an operational way to verify whether two non-degenerate
bipartite mixed states $\rho$ and $\rho^\prime$ are equivalent or
not under local unitary transformations. One only needs to
calculate the matrices $X$ and $Y$ in (\ref{xy}) by calculating
their orthonormal eigenvectors, and check if the rank of the
realigned matrix \cite{realignment} $\tilde{V}$ of $V=XD\,Y^\dag$ could be one,
because it is verified that a matrix $V$ can
be expressed as the tensor product of two matrices
$V_1$ and $V_2$, $V=V_1\otimes V_2$,  if  and  only  if (cf, e.g. \cite{kropro})
$\tilde{V}=vec(V_1)vec(V_2)^t$.
Moreover \cite{tri}, for an $MN\times MN$ unitary matrix $V$,
if $V$ is a unitary decomposable matrix, then the rank of
$\tilde{V}$ is one,  $r(\tilde{V})=1$.
Conversely if $r(\tilde{V})=1$, there exists an $M\times M$ matrix
$U_1$ and an $N\times N$ matrix $U_2$, such that $U=U_1\otimes U_2$ and
$U_1U_1^{\dag}=U_1^{\dag}U_1=k^{-1}I_M$, $U_2U_2^{\dag}=U_2^{\dag}U_2=kI_N$,
where $I_N$ (resp. $I_M$) denotes the $N\times N$ (resp. $M\times M$)
identity matrix, $k>0$, and $V$ is a unitary tensor decomposable matrix.
Therefore if $rank(\tilde{V})=1$, one gets $U_1$
and $U_2$ such that $V=U_1\otimes U_2$, and $\rho$,
$\rho^\prime$ are equivalent under local unitary transformations.

In \cite{yu-eq} the equivalence of bipartite quantum mixed states under local
unitary transformations has been investigated by introducing representation
classes from a geometrical approach.
A mixed state $\rho$ in $H_1\otimes H_2$ with rank $l$ has the spectral decomposition,
\be\label{rho}
\rho=\lambda_{1}|e_{1}\rangle \langle e_{1}|+\cdots +\lambda_{l}
|e_{l}\rangle \langle e_{l}|,
\ee
where $\lambda_{i}$, $i=1,\cdots,l$, are the nonzero eigenvalues of $\rho$,
$|e_{i}\rangle$ are the corresponding eigenvectors associated with $\lambda_{i}$,
$\lambda_{1} \geq \lambda_{2} \geq \cdots \geq \lambda_{l}>0$,
which can be chosen as orthonormal vectors.

Every eigenvector $|e_{i}\rangle$ with Schmidt rank $k_i$ has Schmidt decomposition, namely
there exist orthonormal vectors $a_{j}^{i}$ and $b_{j}^{i}$
of ${H}_{1}$ and ${H}_{2}$ respectively, $j=1,\cdots,
k_{i}$, such that
$$|e_{i}\rangle=\mu_{i}^{1}a_{1}^{i}\otimes b_{1}^{i}+\cdots
+\mu_{i}^{k_{i}}a_{k_{i}}^{i}\otimes b_{k_{i}}^{i},\ \ k_{i} \leq N,
\ \ i=1,\cdots,l,$$
where $\mu_{i}^{j}$, $j=1,\cdots, k_{i}$, are
so called Schmidt coefficients satisfying
$(\mu_{i}^{1})^{2}+ \cdots +(\mu_{i}^{k_{i}})^{2}=1$, $\mu_{i}^{1}\geq \mu_{i}^{2}\geq \cdots \geq
\mu_{i}^{k_{i}}>0$.

Extend the set of $k_1$ orthonormal vectors $a_{1}^{1}, a_{2}^{1}, \cdots, a_{k_{1}}^{1}$ to be an
orthonormal basis of ${H}_{1}$, $\{a_{1}, a_{2}, \cdots , a_{k_{1}}, \cdots , a_{M}\}$,
and $b_{1}^{1}, b_{2}^{1},\cdots, b_{k_{1}}^{1}$ to an orthonormal basis of ${H}_{2}$,
$\{b_{1}, b_{2}, \cdots , b_{k_{1}}, \cdots , b_{N}\}$.
Therefore the vectors $a_{j}^{i}$ and $b_{j}^{i}$, $j=1,\cdots, k_{i}$, can be
represented according to the two bases respectively,
\be\label{XY}
(a_{1}^{i}, a_{2}^{i}, \cdots, a_{k_{i}}^{i})=(a_{1}, a_{2}, \cdots ,
a_{M})X_{i},\ \ \ (b_{1}^{i}, b_{2}^{i}, \cdots,
b_{k_{i}}^{i})=(b_{1}, b_{2}, \cdots , b_{N})Y_{i},
\ee
for some $M\times k_{i}$ matrix $X_{i}$ and $N\times k_{i}$ matrix $Y_{i}$. Denote
$r(\rho)_{i}=(\lambda_{i}, \mu_{i}^{1},\cdots,
\mu_{i}^{k_{i}},X_{i},Y_{i})$, $i=1,...,l$.
We say that
\be\label{rc}
r(\rho)=(r(\rho)_{1}, \cdots,r(\rho)_{l})
\ee
is  a representation of the
mixed state $\rho$. We call the set of all the representations of $\rho$
the representation class of $\rho$, denoted by ${\cal R}(\rho)$.
Then two mixed states $\rho$ and ${\rho}^\prime$ of
bipartite quantum systems are equivalent under local unitary transformations
if and only if they have the same
representation class, i.e. ${\cal R}(\rho)={\cal R}(\tilde{\rho}^\prime).$

In particular for the two-qubit systems, $M=N=2$,
generally a mixed state $\rho$ has four different
eigenvalues $\lambda_{i}, (i=1,\cdots,4)$. Here as the trace of $\rho$ is
one, only three eigenvalues are independent. Let
$|e_{i}\rangle$, $i=1,\cdots,4$, be the corresponding orthonormal
eigenvectors. Since $|e_{4}\rangle$ is determined by other three
eigenvectors up to a scale $e^{i\theta}$, we only need to take into account
three eigenvectors. Every eigenvector of these three
$|e_{i}\rangle$, $i=1,2,3$, can have at most Schmidt rank two.
But only one of the Schmidt coefficients $\mu_{i}^{1}$, $\mu_{i}^{2}$
of $|e_{i}\rangle$, $i=1,2,3$, is independent. Therefore only three
eigenvalues and three Schmidt coefficients (all together 6 quantities) are
free. The matrices $X_{1}$ and $Y_{1}$ are unit matrices of
order $2$. While $X_{2},Y_{2}$, $X_{3},Y_{3}$ are unitary matrices of
order $2$, taking the following form
$$
\left(\begin{array}{cccc} re^{i
\alpha_{1}} &
-\sqrt{1-r^{2}}e^{-i \alpha_{2}}e^{i\alpha_{3}}\\
\sqrt{1-r^{2}}e^{i\alpha_{2}} & re^{-i \alpha_{1}}e^{i\alpha_{3}}
\end{array}\right),
$$
where $r>0$, $\alpha_1,\alpha_3,\alpha_3\in \Rb$. That is,
every matrix has four free quantities. Since $|e_{i}\rangle$,
$i=1,2,3$, are perpendicular to each other, and $|e_{2}\rangle$ and
$|e_{3}\rangle$ are determined up to a phase factor $e^{i\theta}$, there are only
$6$ free parameters left. Therefore we only need at most 12
invariants to check the local equivalence for two-qubit bipartite
quantum systems, which is different from \cite{Makhlin} where 18 invariants are needed.

We see that different approaches give rise to different results.
Usually it is not easy to compare these results. The advantages
of the results depend on detailed classes of quantum states under investigation.
In fact to deal with the equivalence problem $\rho_1 = U_1\otimes \cdots \otimes U_n\,\, \rho_2
U_1^\dagger\otimes \cdots \otimes U_n^\dagger$ is equivalent to
find the standard form of a density matrix $\rho$ under the operations
$U_1\otimes \cdots \otimes U_n\, \rho\, U_1^\dagger\otimes \cdots \otimes U_n^\dagger$.
In the following we investigate the equivalence of quantum states according
to their standard form  under local unitary transformations.
In particular we consider the Schmidt-correlated (SC) states.
We derive the detailed standard form for two-qubit SC states.

Schmidt-correlated (SC) states $\rho = \sum_{m,n=0 }^{N} c_{mn} |m
\cdots m \rangle \langle n \cdots n|$ with $\sum_{m=0}^{N} c_{mm}=1$
is a special class of mixed states \cite{Rains1,ming}. For any local
quantum measurement related to the SC states, the result is same
whichever party performs the measurement. Bipartite SC states
naturally appear in a system dynamics with additive integrals of
motion \cite{Khasin06}. Hence, these states form an important class
of mixed states from a quantum dynamical perspective. From another
point of view, SC states are a natural generalization of pure state.
Therefore we expect SC states have some elegant properties as pure
states. For example, it has been proven that SC state is separable if
and only if it is positive under partial transposition \cite{ming}.

Let ${\cal{A}}$ denote the set of all two-qubit SC
states ${\cal{A}}=\{
c_1|00\rangle\langle00|+c_2|00\rangle\langle11|+c_2^*|11\rangle\langle00|+c_4|11\rangle\langle11|;~
c_1,\,c_4\geq0,\, c_1c_4\geq |c_2|^2,\, c_1+c_4=1\}$. We consider
the local unitary equivalence of two-qubit SC states. We show two states
in ${\cal{A}}$ are LU equivalent if and only if their standard forms
given below coincide. Comparing with the results from other approaches,
our results  are more direct and simple for the SC two-qubit states.

\begin{theorem}\label{th1}
Under local unitary transformation any two-qubit SC state can be transformed into the form
\begin{eqnarray}\label{sc rho}
\rho=\lambda_1|00\rangle\langle00|+\lambda_2|00\rangle\langle11|
+\lambda_2|11\rangle\langle00|+\lambda_4|11\rangle\langle11|
\end{eqnarray}
with non-negative coefficients $\lambda_1, \lambda_2, \lambda_4$ such that
$\lambda_1\lambda_4\geq\lambda_2^2$ and $\lambda_1\geq \lambda_4$.
We call (\ref{sc rho}) the standard form of two-qubit SC states.
\end{theorem}

Proof. For arbitrary mixed state
$\rho=c_1|00\rangle\langle00|+c_2|00\rangle\langle11|+c_2^*|11\rangle\langle00|+c_4|11\rangle\langle11|$
in $\cal{A}$, suppose $c_2=\lambda_2 e^{i\theta}$ with
$\lambda_2\geq 0$. If $c_1\geq c_4$, then set $\lambda_1=c_1$,
$\lambda_4=c_4$. Consequently this state will be transformed into
the standard form under the action of $U\otimes I$ with
$U=e^{-i\theta}|0\rangle \langle 0| + |1\rangle \langle 1|$. If
$c_1< c_4$, then set $\lambda_1=c_4$, $\lambda_4=c_1$. As a result,
this state can be converted into the standard form by $U\otimes U$
with $U=e^{-i\frac{\theta}{2}}|1\rangle \langle 0| + |0\rangle
\langle 1|$. \qed

In fact, we can generalize the above theorem to high dimensional
multipartite case. That is, any SC state can be transformed into
$\rho^\prime = \sum_{m,n=0 }^{N} \lambda_{mn} |m \cdots m \rangle
\langle n \cdots n|$ with non-negative coefficients $\lambda_{mn}$
satisfying $\lambda_{00}\geq \lambda_{11}\geq \cdots \geq
\lambda_{NN}$ and $\lambda_{mm} \lambda_{nn} \geq \lambda_{mn}$.

We now prove that any states in
$\cal{A}$ are LU equivalent if and only if their standard forms coincide.

\begin{theorem}\label{th2}
For arbitrary mixed state
$\rho=c_1|00\rangle\langle00|+c_2|00\rangle\langle11|+c_2^*|11\rangle\langle00|+c_4|11\rangle\langle11|
\in \cal{A}$, $\rho^\prime \in \cal{A}$ is LU equivalent to $\rho$
if and only if
$\rho^\prime=c_1|00\rangle\langle00|+c_2e^{i\delta}|00\rangle\langle11|
+c_2^*e^{-i\delta}|11\rangle\langle00|+c_4|11\rangle\langle11|$
or
$\rho^\prime=c_4|00\rangle\langle00|+c_2e^{i\delta}|00\rangle\langle11|
+c_2^*e^{-i\delta}|11\rangle\langle00|+c_1|11\rangle\langle11|$
with arbitrary real number $\delta$.
\end{theorem}

Proof. Let $U_1=\left(
   \begin{array}{cc}
   a_1 & -a_2 \\
   a_2^* & a_1^*
   \end{array}\right)$ and $U_2=\left(
   \begin{array}{cc}
   b_1 & -b_2 \\
   b_2^* & b_1^*
   \end{array}\right)$ be unitary matrices with
\be\label{unicon}
|a_1|^2+|a_2|^2=1,~~ |b_1|^2+|b_2|^2=1.
\ee
Under the transformations of $U_1\otimes U_2$, the mixed two-qubit SC state
$\rho=c_1|00\rangle\langle00|+c_2|00\rangle\langle11|+c_2^*|11\rangle\langle00|+c_4|11\rangle\langle11|
=\left(
   \begin{array}{cccc}
   c_1 & 0 & 0 & c_2^*\\
   0 & 0 & 0 & 0 \\
   0 & 0 & 0 & 0 \\
  c_2 & 0 & 0 & c_4
   \end{array}\right)$
becomes $\rho^\prime = U_1\otimes U_2 \rho U_1^\dagger\otimes
U_2^\dagger$ with matrix entries:
\begin{eqnarray}
\rho^\prime_{11}=(c_1a_1b_1+c_2^*a_2b_2)a_1^*b_1^* +
(c_2a_1b_1+c_4a_2b_2)a_2^*b_2^* \label{11} ,\\
\rho^\prime_{12}=(c_1a_1b_1+c_2^*a_2b_2)a_1^*b_2 -
(c_2a_1b_1+c_4a_2b_2)a_2^*b_1 ,\\
\rho^\prime_{13}=(c_1a_1b_1+c_2^*a_2b_2)a_2b_1^* -
(c_2a_1b_1+c_4a_2b_2)a_1b_2^* ,\\
\rho^\prime_{14}=(c_1a_1b_1+c_2^*a_2b_2)a_2b_2 +
(c_2a_1b_1+c_4a_2b_2)a_1b_1 \label{14},\\
\rho^\prime_{21}=(c_1a_1b_2^*-c_2^*a_2b_1^*)a_1^*b_1^* +
(c_2a_1b_2^*-c_4a_2b_1^*)a_2^*b_2^* ,\\
\rho^\prime_{22}=(c_1a_1b_2^*-c_2^*a_2b_1^*)a_1^*b_2 -
(c_2a_1b_2^*-c_4a_2b_1^*)a_2^*b_1 ,\\
\rho^\prime_{23}=(c_1a_1b_2^*-c_2^*a_2b_1^*)a_2b_1^* -
(c_2a_1b_2^*-c_4a_2b_1^*)a_1b_2^* ,\\
\rho^\prime_{24}=(c_1a_1b_2^*-c_2^*a_2b_1^*)a_2b_2 +
(c_2a_1b_2^*-c_4a_2b_1^*)a_1b_1 ,\\
\rho^\prime_{31}=(c_1a_2^*b_1-c_2^*a_1^*b_2)a_1^*b_1^* +
(c_2a_2^*b_1-c_4a_1^*b_2)a_2^*b_2^* ,\\
\rho^\prime_{32}=(c_1a_2^*b_1-c_2^*a_1^*b_2)a_1^*b_2 -
(c_2a_2^*b_1-c_4a_1^*b_2)a_2^*b_1 ,\\
\rho^\prime_{33}=(c_1a_2^*b_1-c_2^*a_1^*b_2)a_2b_1^* -
(c_2a_2^*b_1-c_4a_1^*b_2)a_1b_2^* ,\\
\rho^\prime_{34}=(c_1a_2^*b_1-c_2^*a_1^*b_2)a_2b_2 +
(c_2a_2^*b_1-c_4a_1^*b_2)a_1b_1 ,\\
\rho^\prime_{41}=(c_1a_2^*b_2^*+c_2^*a_1^*b_1^*)a_1^*b_1^* +
(c_2a_2^*b_2^*+c_4a_1^*b_1^*)a_2^*b_2^* \label{41},\\
\rho^\prime_{42}=(c_1a_2^*b_2^*+c_2^*a_1^*b_1^*)a_1^*b_2-
(c_2a_2^*b_2^*+c_4a_1^*b_1^*)a_2^*b_1 ,\\
\rho^\prime_{43}=(c_1a_2^*b_2^*+c_2^*a_1^*b_1^*)a_2b_1^* -
(c_2a_2^*b_2^*+c_4a_1^*b_1^*)a_1b_2^* ,\\
\rho^\prime_{44}=(c_1a_2^*b_2^*+c_2^*a_1^*b_1^*)a_2b_2 +
(c_2a_2^*b_2^*+c_4a_1^*b_1^*)a_1b_1. \label{44}
\end{eqnarray}
Because $\rho^\prime \in \cal{A}$, the entries are all zeros except
$\rho^\prime_{11}$, $\rho^\prime_{14}$,$\rho^\prime_{41}$ and $\rho^\prime_{44}$.
Taking into account the Hermitian property of the density matrix we have the following
relations:
\begin{eqnarray}
c_1|a_1|^2b_1b_2+c_2^*a_1^*a_2b_2^2-c_2a_1a_2^*b_1^2-c_4|a_2|^2b_1b_2=0\label{12},\\
c_1a_1a_2|b_1|^2+c_2^*a_2^2b_1^*b_2-c_2a_1^2b_1b_2^*-c_4a_1a_2|b_2|^2=0\label{13},\\
c_1|a_1|^2|b_2|^2-c_2^*a_1^*a_2b_1^*b_2-c_2a_1a_2^*b_1b_2^*+c_4|a_2|^2|b_1|^2=0\label{22},\\
c_1a_1a_2b_1^*b_2^*-c_2^*a_2^2b_1^{*2}-c_2a_1^2b_2^{*2}+c_4a_1a_2b_1^*b_2^*=0\label{23},\\
c_1a_1a_2|b_2|^2-c_2^*a_2^2b_1^*b_2+c_2a_1^2b_1b_2^*-c_4a_1a_2|b_1|^2=0\label{24},\\
c_1|a_2|^2|b_1|^2-c_2^*a_1^*a_2b_1^*b_2-c_2a_1a_2^*b_1b_2^*+c_4|a_1|^2|b_2|^2=0\label{33},\\
c_1|a_2|^2b_1b_2-c_2^*a_1^*a_2b_2^2+c_2a_1a_2^*b_1^2-c_4|a_1|^2b_1b_2=0.\label{34}
\end{eqnarray}
Adding Eq. (\ref{12}) and Eq. (\ref{34}), Eq. (\ref{13}) and Eq. (\ref{24}) we have
$$
(c_1-c_4)b_1b_2=0,~~~(c_1-c_4)a_1a_2=0.
$$
We analyze below for different cases:

{\em Case} 1. $c_1\neq c_4$.

In this case we have $b_1b_2=0$ and $a_1a_2=0$. Utilizing these two
conditions, Eqs. (\ref{22}), (\ref{23}) and (\ref{33}) become
\begin{eqnarray}
c_1|a_1|^2|b_2|^2+c_4|a_2|^2|b_1|^2=0\label{new22},\\
c_2^*a_2^2b_1^{*2}+c_2a_1^2b_2^{*2}=0\label{new23},\\
c_1|a_2|^2|b_1|^2+c_4|a_1|^2|b_2|^2=0\label{new33}.
\end{eqnarray}

Since $c_1\geq0$ and $c_4\geq0$, we have $a_1b_2=0$ and $a_2b_1=0$
by Eq. (\ref{new22}) or Eq. (\ref{new33}).

(1.1) If $a_1=b_1=0$, then $a_2=e^{i\theta}$ and $b_2=e^{i\phi}$. In this case we get,
\begin{eqnarray}\label{1.1}
\rho^\prime&=&\left(
   \begin{array}{cccc}
   c_4 & 0 & 0 & c_2e^{i2\gamma}\\
   0 & 0 & 0 & 0 \\
   0 & 0 & 0 & 0 \\
  c_2^*e^{-i2\gamma} & 0 & 0 & c_1
   \end{array}\right)\nonumber\\[1mm]
   &=&c_4|00\rangle\langle00|+c_2e^{i\gamma}|00\rangle\langle11|+c_2^*e^{-i\gamma}|11\rangle\langle00|
   +c_1|11\rangle\langle11|
   \end{eqnarray}
with $\theta+\phi=\gamma$.

(1.2) If $a_2=b_2=0$, then $a_1=e^{i\theta}$ and $b_1=e^{i\phi}$. We obtain
\begin{eqnarray}\label{1.2}
\rho^\prime&=&\left(
   \begin{array}{cccc}
   c_1 & 0 & 0 & c_2e^{i2\gamma}\\
   0 & 0 & 0 & 0 \\
   0 & 0 & 0 & 0 \\
  c_2^*e^{-i2\gamma} & 0 & 0 & c_4
   \end{array}\right)\nonumber\\[1mm]
   &=&c_1|00\rangle\langle00|+c_2e^{i\gamma}|00\rangle\langle11|
   +c_2^*e^{-i\gamma}|11\rangle\langle00|+c_4|11\rangle\langle11|
   \end{eqnarray}
with $\theta+\phi=\gamma$.

Therefore we get that if $\rho^\prime \in \cal{A}$  is LU equivalent to
$\rho$, then it has the form (\ref{1.1}) or (\ref{1.2}).

{\em Case} 2. $c_1=c_4=\frac{1}{2}$.

In this case Eqs. (\ref{12}) and (\ref{34}), (\ref{13}) and
(\ref{24}), (\ref{22}) and (\ref{33}) are equivalent respectively.
We only need to consider equations (\ref{12}), (\ref{13}),
(\ref{22}) and (\ref{23}).

$ {\rm Eq.} (\ref{13})\times b_2^* -  {\rm Eq.}(\ref{23})\times b_1$
and $ {\rm Eq.}(\ref{12})\times b_2^* -  {\rm Eq.}(\ref{22})\times
b_1$ give rise to
\begin{eqnarray}
c_2^*a_2^2b_1^*=c_4a_1a_2b_2^*\label{a},\\
c_2^*a_1^*a_2b_2=c_4|a_2|^2b_1\label{b}
\end{eqnarray}
respectively. Then ${\rm Eq.}(\ref{a})\times a_1^*b_2 - {\rm
Eq.}(\ref{b})\times a_2b_1^*$ gives rise to
\begin{eqnarray}
a_2(|a_1b_2|^2-|a_2b_1|^2)=0.
\end{eqnarray}

(2.1) If $a_2=0$, then $b_2=0$ by Eq. (\ref{23}). In this situation,
the mixed state $\rho^\prime$ has the form as Eq. (\ref{1.2}).

(2.2) Assume $|a_1b_2|=|a_2b_1|$.

(2.2.1) If $|a_1b_2|=|a_2b_1|=0$, then we need only to consider
$a_1=b_1=0$, which shows that $\rho^\prime$ has the form as Eq. (\ref{1.1}).

(2.2.2) If $|a_1b_2|=|a_2b_1|\neq0$, then we get $|c_2|=\frac{1}{2}$
by taking module on Eq. (\ref{a}). Suppose
$c_2=\frac{1}{2}e^{i\alpha}$, one has
\begin{eqnarray}\label{d}
e^{-i\alpha}a_2b_1^*=a_1b_2^*
\end{eqnarray}
by substituting the value $c_2$ to Eq. (\ref{a}). From Eq.
(\ref{d}) and the value $c_2$, Eq. (\ref{23}) becomes
\begin{eqnarray}
2-e^{-i2\alpha}-e^{i2\alpha}=0,
\end{eqnarray}
which indicates $\alpha=0$, $c_2=\frac{1}{2}$ and
\begin{eqnarray}
a_2b_1^*=a_1b_2^*.
\end{eqnarray}
This equation can also be expressed as
\begin{eqnarray}\label{e}
\frac{a_2}{b_2^*}=\frac{a_1}{b_1^*}=e^{i\beta}
\end{eqnarray}
due to Eq. (\ref{unicon}). Using Eq. (\ref{e}) and
$c_1=c_2=c_4=\frac{1}{2}$, one can simplify Eqs. (\ref{11}),
(\ref{14}), (\ref{41}) and (\ref{44}). At last we get
\begin{eqnarray}
\rho^\prime&=&\left(
   \begin{array}{cccc}
   \frac{1}{2} & 0 & 0 & \frac{1}{2}e^{i2\beta}\\
   0 & 0 & 0 & 0 \\
   0 & 0 & 0 & 0 \\
  \frac{1}{2}e^{-i2\beta} & 0 & 0 & \frac{1}{2}
   \end{array}\right)\nonumber\\[1mm]
   &=&\frac{1}{2}(|00\rangle\langle00|+e^{i\beta}|00\rangle\langle11|
   +e^{-i\beta}|11\rangle\langle00|+|11\rangle\langle11|).
\end{eqnarray}

Therefore the theorem holds true also for the case $c_1= c_4$. \qed

From theorem \ref{th1} and \ref{th2} we have the necessary and sufficient condition of the LU
equivalence of mixed state in $\cal{A}$:

\begin{theorem}
Two SC states $\cal{A}$ are LU equivalent if and only if their
standard forms coincide.
\end{theorem}

In particular for the pure states in $\cal{A}$, i.e. all
two-qubit pure states having the same Schmidt basis, we have
get the corresponding necessary and sufficient condition:

\begin{cor}
Pure state $|\phi\rangle=b_0|00\rangle + b_1|11\rangle$ is LU equivalent to
$|\psi\rangle = a_0|00\rangle + a_1|11\rangle$, $b_0\geq b_1 \geq 0$, $b_0\geq b_1 \geq 0$, if and only if $b_0 = a_0$ and $b_1= a_1$.
\end{cor}

In the following we study the correlations of two-qubit
states in ${\cal{A}}$. Correlations also characterize some kinds of
properties of quantum states \cite{A. Datta,B. P. Lanyon}. Generally a quantum state
contains both classical and quantum correlations. In \cite{H.
Ollivier, S. Luo} the classical correlation is defined in terms of
measurement-based conditional density operators, and the quantum
correlation is defined as the difference of quantum mutual information and
classical correlation. While in \cite{K. Modi}
classical correlation and quantum correlation are quantified by using relative entropy
\cite{V. Vedral1997, V. Vedral1998} as a distance measure. Both
definitions have their own advantages, but in both cases the relating calculations
are quite difficult. Recently the quantum discord for two-qubit $X$ states has been
computed in terms of measurement-based conditional density operators \cite{xstate}.

The classical correlation of a bipartite quantum state can be defined in
terms of measurement-based density operator \cite{H. Ollivier, S.
Luo}. Let ${B_k}$ be projectors performed locally on party $B$,
then the quantum state, conditioned on the measurement outcome
labeled by $k$, changes to
\begin{eqnarray}
\rho_k=\frac{1}{p_k} (I\otimes B_k) \rho (I\otimes B_k)
\end{eqnarray}
with probability $p_k= tr (I\otimes B_k) \rho (I\otimes B_k)$.
Clearly, $\rho_k$ may be considered as a conditional density
operator. With this conditional density operator, the quantum
conditional entropy with respect to this measurement is given by
\begin{eqnarray}
S(\rho|\{B_k\})=\sum_k p_k S(\rho_k)
\end{eqnarray}
and furthermore the associated quantum mutual information of this
measurement is defined as
\begin{eqnarray}
{\cal{I}} (\rho|\{B_k\})=S(\rho_A)- S(\rho|\{B_k\}).
\end{eqnarray}
A measure of the resulting classical correlation is provided by
\begin{eqnarray}
{\cal{C}}_M(\rho) \equiv \sup_{\{B_k\}} {\cal{I}} (\rho|\{B_k\}).
\end{eqnarray}
The difference of mutual information ${\cal{I}}(\rho)$ and classical
correlation ${\cal{C}}_M (\rho)$ is called quantum discord
\begin{eqnarray}
{\mathcal {D}}_M (\rho)= {\cal{I}}(\rho)- {\cal{C}}_M (\rho),
\end{eqnarray}
which is a kind of measure of quantum correlation. Here the quantum mutual
information is
\begin{eqnarray}
{\cal{I}}(\rho_{AB})=S(\rho_A) + S(\rho_B) - S(\rho)
\end{eqnarray}
with $S(\rho)=-tr(\rho \log \rho)$ and $\rho_{A(B)}=tr_{B(A)}\rho$.

Consider the correlations in two-qubit SC states
$\rho=c_1|00\rangle\langle00|+c_2|00\rangle\langle11|+c_2^*|11\rangle\langle00|+c_4|11\rangle\langle11|$
according to the above definition. Let $\{ \Pi_k = |k\rangle \langle
k|: k=0,1\}$ be the local measurement for party $B$, then any
projector for party $B$ can be written as $\{ B_k =V\Pi_k V^\dagger:
k=0,1 \}$ for some unitary $V\in SU(2)$. While any unitary $V$ can
be written as $V= tI+i\vec{y}\cdot\vec{\sigma}$ with $t\in \Rb$,
$\vec{y}=(y_1, y_2, y_3)\in \Rb^3$, and $t^2 + y_1^2 +y_2^2 +y_3^2
=1$. By employing the
relations $\Pi_0 \sigma_3 \Pi_0 = \Pi_0$, $\Pi_1 \sigma_3 \Pi_1 =
-\Pi_1$, $\Pi_j \sigma_k \Pi_j = 0$ for $j=0,1$, $k=1,2$ and
$V^\dagger \sigma_3 V =2(ty_2 +y_1y_3) \sigma_1 + 2(-t y_1 + y_2
y_3)\sigma_2 +(t^2 +y_3^2 -y_1^2 -y_2^2)\sigma_3$,
\begin{eqnarray}
p_0&=& \frac{1}{2}(1+(c_1-c_4)x),\\\nonumber
p_1&=&\frac{1}{2}(1+(c_4-c_1)x),\\\nonumber \rho_0 &=&
\frac{1}{2}(I+\frac{(c_1-c_4)+ x}{1+(c_1-c_4)x} \sigma_3) \otimes
V\Pi_0V^\dagger, \\\nonumber \rho_1 &=&
\frac{1}{2}(I+\frac{(c_1-c_4)- x}{1+(c_4-c_1)x} \sigma_3) \otimes
V\Pi_1V^\dagger, \\\nonumber
\end{eqnarray}
with $x=t^2 +y_3^2 -y_1^2 -y_2^2$. By straightforward calculations we get
$$
\ba{rcl}
\inf_{\{B_k\}} \sum_k p_k S(\rho_k) &=& -(\sup_{\{B_k\}} (c_1(1+x) \log
\frac{c_1(1+x)}{1+(c_1-c_4)x}
+ c_4(1-x) \log\frac{c_4(1-x)}{1+(c_1-c_4)x}\\[2mm]
&& + c_1(1-x) \log
\frac{c_1(1-x)}{1+(c_4-c_1)x} + c_4(1+x) \log
\frac{c_4(1+x)}{1+(c_4-c_1)x}))\\[3mm]
&=& -(c_1\log c_1 + c_4 \log c_4).
\ea
$$

Therefore the classical correlation in $\rho$ is ${\cal{C}}_M (\rho)=
S(\rho_A) - \inf_{\{B_k\}} \sum_k p_k S(\rho_k) = 0$, which shows that the
total correlation in two-qubit SC states is just the quantum
correlation. Hence we get that the discord of two-qubit SC state is
$2S(\rho_A) - S(\rho)= -2 ( c_1\log c_1 + c_4\log c_4)-S(\rho)$,
where $S(\rho)= \frac{1+\sqrt{\Delta}}{2}\log
\frac{1+\sqrt{\Delta}}{2} + \frac{1-\sqrt{\Delta}}{2}\log
\frac{1-\sqrt{\Delta}}{2}$ with $\Delta=1-4c_1 c_4 + 4 |c_2|^2$.

Different from the correlations described in terms of
measurement-based density operator, Ref. \cite{K. Modi} measures the
correlations in a given quantum state by using the relative
entropy \cite{V. Vedral1997, V. Vedral1998} as a distance measure.
For a given quantum state $\rho$, the discord is defined as
${\cal{D}}_R=\min_{\chi\in \cal{C}}S(\rho||\chi)$, and the classical
correlation ${\cal{C}}_R=\min_{\pi\in \cal{P}}S(\rho||\pi)$ with
$\cal{C}$ resp. $\cal{P}$ denoting the set of all classical resp. product
states respectively. By direct calculation, one can find that the
closest classical state to two-qubit SC state
$\rho=c_1|00\rangle\langle00|+c_2|00\rangle\langle11|+c_2^*|11\rangle\langle00|+c_4|11\rangle\langle11|$
is $ \chi_0 =c_{1} |00 \rangle \langle 00 |+c_{4} |11 \rangle
\langle 11|$. Hence the discord in $\rho$ is ${\cal{D}}_R=S(\rho ||
\chi_0)=-(c_1\log c_1 + c_4\log c_4)-S(\rho)$.

The classical correlation is given by ${\cal{C}}_R=S(\rho || \pi_0)=-2(c_1^2\log c_1 +
c_4^2\log c_4)-S(\rho)$ with the closest product state to $\rho$ being
$\pi_0= (c_{1} |0\rangle \langle 0|+c_{4} |1\rangle \langle
1|)^{\otimes 2}$. We see that the classical correlation is nonzero for entangled two-qubit SC states $\rho$.
This result is quite different from the correlations defined in terms of
measurement-based density operators, which detects no classical correlation for
two-qubit SC states.

In addition, if one looks at the entanglement in two-qubit SC state
$\rho$ in terms of relative entropy ${\cal{ E}}_R(\rho) =\min _{
\sigma \in {\cal{D}}} S( \rho \parallel \sigma )$,
where ${\cal{D}}$ is the set of separable states, one finds that
${\cal{ E}}_R={\cal{D}}_R$ for two-qubit SC state, i.e. the
entanglement and quantum correlation are the same for such states.

We have presented the sufficient and necessary conditions of
whether two different two-qubit SC states are LU equivalent by
deriving the standard forms. The correlations for two-qubit SC
states are also investigated. It has been shown that the
classical correlation is zero in terms of the definition in
\cite{H. Ollivier, S. Luo}, and nonzero according to the definition in
\cite{K. Modi}.


\end{document}